\documentstyle[aps,epsfig,float,twocolumn,prl]{revtex}
\newcommand{\be}{\begin{equation}}
\newcommand{\ee}{\end{equation}}

\begin{document}
\twocolumn[\hsize\textwidth\columnwidth\hsize\csname @twocolumnfalse\endcsname
\draft
\title{Entropic Tightening of Vibrated Chains}
\author{M. B. Hastings$^{1,2}$,
  Z. A. Daya$^{2,3}$, E. Ben-Naim$^{1,2}$, and R. E. Ecke$^{2,3}$}
\address{$^1$Theoretical Division, $^2$Center for Nonlinear Studies,
$^3$Condensed Matter \& Thermal Physics Group\\
Los Alamos National Laboratory, Los Alamos, NM 87545}
\date{October 15, 2001}
\maketitle
\begin{abstract}
We investigate experimentally the distribution of configurations of a
ring with an elementary topological constraint, a ``figure-8''
twist.  Using vibrated granular chains, which permit controlled
preparation and direct observation of such a constraint, we show that
configurations where one of the loops is tight and the second is large
are strongly preferred.  This agrees with recent predictions for
equilibrium properties of topologically-constrained polymers.  However, the
dynamics of the tightening process weakly violate detailed balance, a
signature of the nonequilibrium nature of this system.
\vskip1mm {PACS:} 05.40.-a, 81.05.Rm, 83.10Nn \vskip1mm
\end{abstract}
]

Topological constraints such as knots and entanglements constantly
form and relax in polymer systems and in biomolecules such as
DNA\cite{n,fw,sw,sw1,q,bgwb}.  These constraints increase relaxation
time scales and additionally restrict the phase space accessible by
the macromolecule. Whereas the role entanglements play in chain dynamics
is well appreciated \cite{dg,de}, even more basic questions concerning
effects of topological constraints on static properties such as the
chain structure remain largely unanswered.  Recent theoretical
studies predict that in equilibrium, a knotted polymer will generally
favor configurations where the knot is ``tight", {\it i.e.}, localized to a
small region of the chain\cite{kovds,g}.  Numerical simulations and
scaling analysis support this prediction \cite{mhdkk}, but direct
experimental tests are difficult \cite{qbc}.  In this study, we
examine this interesting prediction experimentally using vertically
vibrated granular chains.

Granular chains consist of spherical beads connected by rods, and
their backbone enforces the same geometrical constraints as in a
polymer system. A vibrating plate supplies the system with energy,
balancing the energy dissipation due to inelastic collisions
experienced by beads \cite{k,jnb,ou,mus}. This polymer system is well
suited for studying topological constraints as it allows control of
the chain size and the constraint type, as well as direct observations
of the chain conformation in contrast with traditional polymer systems.
Recent studies have successfully used vibrated chains to
study diffusive relaxation \cite{bdve} and spontaneous formation
\cite{bsew} of knots. 
 
We considered the simplest possible topology, 
a ``figure-8'': a once-twisted ring
consisting of two loops, separated by a single crossing point, which
functions as the topological constraint (see Fig.~1).  
Under appropriate vibration amplitude, the system is effectively 
two-dimensional and the crossing
point hops along the chain without flipping open the figure-8.   
Surprisingly, despite the highly nonequilibrium drive applied to the
system, we observed strong entropic tightening.  
The microscopic degrees of freedom, the beads,
experience periodic drive and dissipative collisions with the
plate, rods, and other beads, as well as frictional forces.
Remarkably the macroscopic observable, the loop size,
obeys a statistical mechanics.
Detailed balance is only
weakly violated and the empirical
loop-size distribution is close to that conjectured on
entropic grounds. 
This provides an opportunity for studying the role of entropy in
nonequilibrium systems.
We now review the equilibrium result for the loop-size distribution.

\begin{figure}[!t]
\begin{center}
\leavevmode \epsfig{figure=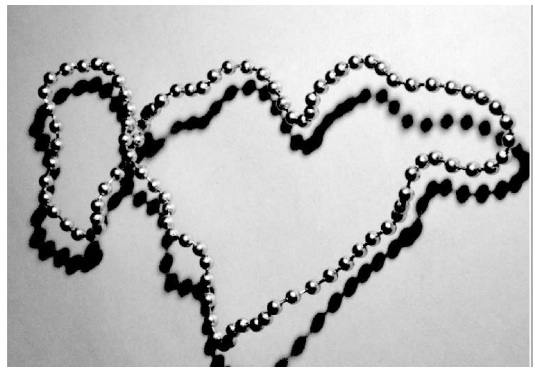,height=10cm,angle=90,scale=.7}
\end{center}
\caption{A vibrated ring with a figure-8 twist.}
\end{figure}

Tightening of knots in polymers can be understood by considering
the simplest case, in which electrostatic interactions are
perfectly screened. Suppose a knot in the polymer can be considered as
located at a point, with several loops of lengths $N_1,N_2,...$
projecting from the knot.  Ignoring hard-core interactions, each loop
has a linear size of order $N_i^{1/2}$, so that the number of
configurations in which the chain returns to itself and forms a loop
is proportional to $e^{cN} \prod_i N_i^{-d/2}$, where $d=2$ is the
spatial dimension, and $c$ is a constant.  Therefore, assuming
all microscopic configurations are equally likely, the 
probability of having a given $N_1,N_2,\cdots$ is 
maximized when $N_1 \approx N$ and $N_2,\cdots$
are all as short as possible. That is, the knot spontaneously tightens
due to entropic effects.

A ring with a figure-8 constraint is natural for studying this
effect as it simply contains two loops of sizes $N_1$,
$N_2$ with the overall size $N=N_1+N_2$ fixed. The argument above
predicts a power law divergence in the loop-size probability distribution 
\begin{equation}
\label{powerlaw} 
\rho(n)\propto n^{-\alpha}
\end{equation}
with $n\equiv N_1,N_2$ for $1 \ll n\ll N$. The exponent
$\alpha=43/16 \approx 2.7$ can be obtained even when excluded volume interactions
are taken into account \cite{mhdkk,ds}.  Although this is obtained
from large $n,N$ asymptotics, an enumeration for smaller sizes only
leads to small corrections to this form.  

Our apparatus consists of an anodized aluminum plate, driven
sinusoidally by an electromechanical vibrator. The bead and
rod chains consist of hollow nickel-plated stainless steel spheres of
radius $1.18 \pm 0.01$~mm connected by thin rigid rods of radius $0.26
\pm 0.01$~mm. The rods constrain both the bending and stretching of
the chain. In particular, the rods must lie within a cone of a
half-angle of roughly $25^\circ$ about the axis of either of the two
adjacent rods, and the separation $b$ between two adjacent beads lies
in the range $0 \leq b \leq 0.94$~mm. The chains were connected
end-to-end to form rings, and then twisted with a single crossing
point thereby forming a figure-8. For the experiments reported
here, the number of beads in the figure-8 ring was between $69$
and $219$, much larger than the tightest possible $8$-bead loop. The
plate was oscillated harmonically at a frequency of $16$~Hz.  

The dynamics of the crossing depend on the rms acceleration of the
plate, $\gamma$ (this dimensionless quantity is in units of the
gravitational acceleration $g$).  For $\gamma\lesssim 1.35$, the
crossing does not move along the chain.  For $\gamma\gtrsim 1.55$, the
vertical motion of the chain is large enough that the number of
crossings in the ring is not fixed: a loop of the figure-8 can
easily flip, untwisting the ring, or creating additional crossings.
We chose $\gamma=1.5$, for which flipping events remained rare,
occurring every roughly $10^4$ oscillation cycles, while the crossing
remained mobile, with approximately 50\% chance of the loop size
changing in a $1/16$-th second cycle.  In Fig.~2 we show the
distribution of change in loop size in one cycle.  The acceleration
was constant and uniform across the plate to better than $1\%$. The
$27.2$~cm plate diameter, corresponding to approximately $115$ beads,
was large enough so that collisions with the sloped acrylic wall were
rare.

Digital images of the chain were obtained to determine the loop
size distribution. Image analysis requires a two step procedure
involving: (i) monomer recognition, and (ii) chain reconstruction.
To obtain the monomer positions, images of resolution $1000 \times1016$
pixels were acquired.  At this resolution, the reflected
light from a bead appears in the images as a bright spot of about
$5 \times 5$ pixels, even though the bead has a diameter of just
over $8$ pixels. The positions of the beads were determined by
fitting the intensity pattern generated by each bead to a
Gaussian, with the peak position taken as the bead 
position. We
estimate the positional accuracy obtained using
this procedure as $0.05$ bead diameters.
\begin{figure}[!t]
\begin{center}
\leavevmode \epsfig{figure=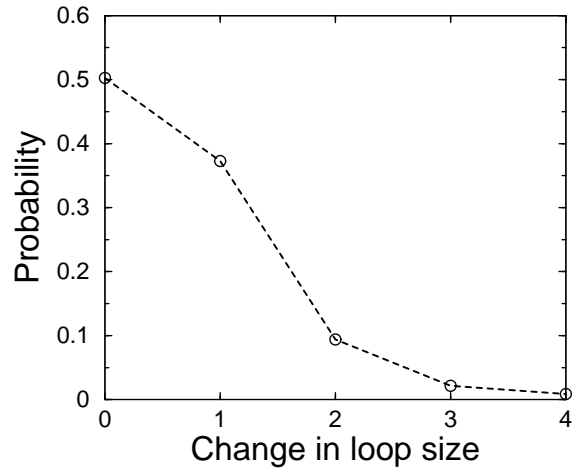,height=8.cm,angle=0,scale=.8}
\end{center}
\caption{Distribution of absolute change in loop size over $1/16$-th second,
showing the rapidly decreasing probability of larger jumps.}
\end{figure}
Given these positions, the order of the monomers along the chain
was determined using an efficient greedy algorithm which requires
only $N^2$ operations for an $N$-bead ring. This algorithm
utilizes the aforementioned geometrical restrictions on stretching
and bending imposed by the rods (given two connected beads, the
third was searched only in a properly restricted neighborhood).
Once the ring was reconstructed, the crossing point was
identified, thereby determining the loop sizes.

We first sampled the loop size at a much slower rate of $0.25$ Hz.
Starting with two equal-size loops at each run for a chain of size
$N=149$, we obtained the loop-size distribution shown as the inset to
Fig.~3.  This distribution has a sharp peak located at the smallest
possible loop size.  Hence, the conjectured tightening is indeed
observed. We repeated the experiments using larger beads, changing the
driving frequency to $13$ Hz, and using chains of length 49, 69, 99,
and 219, as well as using a plate with different
roughness. Further, many sets of images at arbitrary phase with
respect to the driving were analyzed. In all these cases, the same
qualitative loop-size distribution emerged.  There were, however, some
quantitative differences with the peak height varying by about 20\%.
Since the number of hopping events before flipping is of the same
order as $N^2$ ($\sim 10^4$), the flipping introduces a dependence on
the initial conditions, which is more pronounced for the longer
chains. Even worse, flipping events typically occur when the loops are
small, leading to a reduction of the peak height.  Therefore, we used
an alternate method of determining the loop size distribution.

To find the true peak height, {\it with flipping events removed}, we
measured the loop size at a much faster frame rate of $16$ Hz, and
experimentally determined the transition probability $t_{i,j}$ from a
loop of size $i$ to a loop of size $j$. To sample $t_{i,j}$ for all
$i$, the twisted ring was started manually at various equally-spaced
loop sizes, and then allowed to run for 200 cycles to let the chain
equilibrate, after which 200 frames were taken to measure $t_{i,j}$.
By taking 20,000 total frames over 100 separate runs, we obtained an
accuracy of $10$\% on individual $t_{i,j}$.  After equilibration,
correlations between successive transitions were negligible, implying
a Markov process.

Given Markovian dynamics,
it is possible to calculate the steady state probability,
$\rho_i$, of the loop having size $i$, from the $t_{i,j}$ by
performing a Monte Carlo simulation of the transition process.  For a
chain of length 149, we found the distribution shown as the solid line
in Fig.~3.  Compared to that measured directly, this histogram is
characterized by a sharper peak and considerable curvature at the
center because flipping events and dependence on initial conditions
were eliminated, respectively.  The dashed line shows the theoretical
curve $\rho_i\propto \left[ i^{-\alpha} (N-i)^{-\alpha}\right]$, with
$\alpha=43/16$.  The two curves are consistent, although the peak is
more sharply defined in the solid line.  Making quantitative
statements about the relation between the curves would require much
larger chain lengths $N$ to obtain a sufficient scaling region.
Further, the statistical error in $\rho_i$ is most pronounced in the
center of the histogram where $\rho_i$ is small.  Other chain lengths
have similar histograms, in this case using $0.25$ Hz transition
rates.
\begin{figure}[!t]
\begin{center}
\leavevmode \epsfig{figure=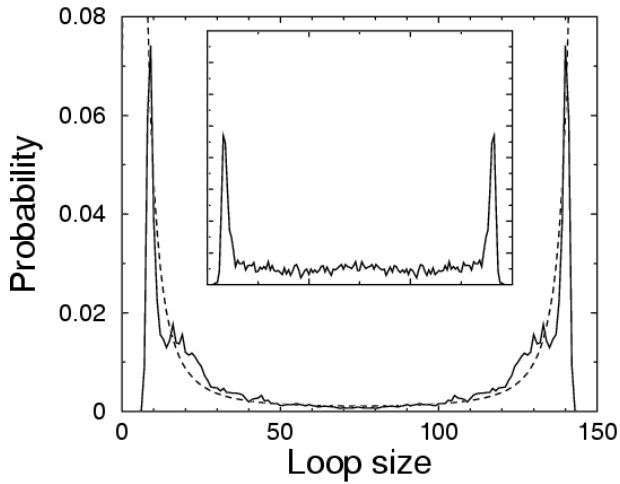,height=8.cm,angle=0,scale=.8}
\end{center}
\caption{Loop-size distribution obtained from transition rates (solid line),
and equilibrium result (dashed line).
Inset: Loop-size distribution from direct observation, with
same scale as main figure.}
\end{figure}

The transition rates enable us to check detailed balance, a sharp test
of the nonequilibrium nature of the system.  For a system in thermal
equilibrium, detailed balance implies a vanishing net flux between any
two microscopic states, namely $\rho_i t_{i,j}=\rho_j t_{j,i}$, or
\hbox{$f_j(i)\equiv \ln{\left[(\rho_i t_{i,i+j})/ (\rho_{i+j}
t_{i+j,i})\right]}=0$}.  Interestingly, we have found that $f_1(i)>0$
and $f_2(i)<0$, namely, short jumps tend to tighten the loop
more than long jumps.  This is shown in Fig.~4, where we plot a moving
average of $f_1(i)$ as the solid line, and a moving average of
$f_2(i)$ as the dashed line.  The average of $f_1(i)$ is
$0.1\pm 0.02$ for $i>100$.
\begin{figure}[!t]
\begin{center}
\leavevmode \epsfig{figure=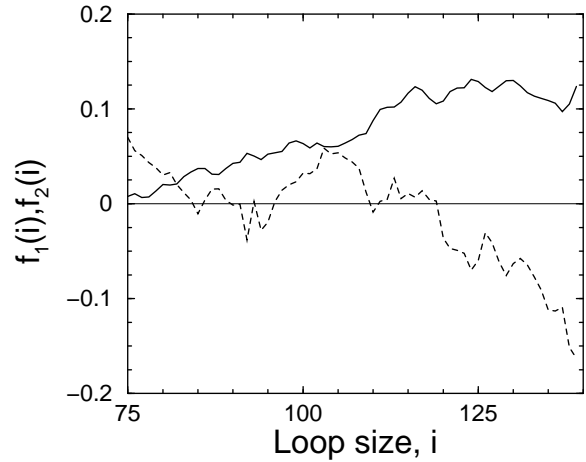,height=8.cm,angle=0,scale=.8}
\end{center}
\caption{Plot of smoothed $f_1(i)$ (solid line), and smoothed
$f_2(i)$ (dashed line).}
\end{figure}

As this is an important point, we have made several additional checks.
We also considered
$f_1(i)+f_1(i+1)-f_2(i+2)$, which measures the flux around a three
point neighborhood.  This $\rho$-independent quantity was positive,
demonstrating violation of detailed balance directly from the
$t_{i,j}$.  Furthermore, we tested detailed balance
using $0.25$ Hz transition rates for other chain lengths, and again
found that short jumps tighten the loop more than long jumps.  As a
final check, we have tested this procedure using surrogate data from
a simulated process which satisfies detailed balance to verify that
the violation is not an artifact of the reconstruction of $\rho$ from
the $t_{i,j}$.

Since it is surprising that an argument based on counting of states
should be relevant in a nonequilibrium system, we now consider a
simplified model.  We show that, even out of equilibrium, there are
entropic tightening forces, despite violation of detailed balance and
possible quantitative changes in the size distribution.  We consider a
chain with linear elastic interactions between neighboring beads and
ignore self-avoidance.  We will first consider the equilibrium case,
with the chain subject to thermal forcing and damping, and then
generalize to athermal drive.  Number the beads from $1$ to $N$, and
label the position of bead $i$ by $\vec x(i)$.  Let the crossing occur
at beads $n_1,n_1+1$ and $n_2,n_2+1$.  We will compute the forces on
the crossing, and from this derive an effective dynamics.

Assume $n_1<n_2$, and consider the loop formed by beads $n_1+2,n_1+3,...,n_2-2$.
This loop exerts a force on bead $n_1+1$ proportional to $\vec x(n_1+2)-
\vec x(n_1+1)$.  This force tries to tighten the loop.
Summing over normal modes of the loop, the
average force of a loop of length $n$ is found to be
$k T(c-1/n)$, with $c>0$.  The fluctuations in this force introduce
an effective noise into the dynamics of $n_1$.  There is also
an effective dissipation: 
if the crossing moves to tighten a given loop, the force exerted by that loop
temporarily increases (both these contributions are determined by
short distance effects).  
We then make an approximation that $n_1,n_2$ can be treated as
particles of mass $M$ subject to the above forces.
The resulting dynamics for $n=n_2-n_1$ when $n<<N$ is, 
\be
\label{cdyn}
M\partial_{t}^2n=-\frac{kT}{n}-\nu \partial_t n+\eta(t),
\ee
where $\nu$ is the dissipation parameter and $\eta$ is the noise.
By the fluctuation-dissipation theorem, noise and
dissipation in (\ref{cdyn}) are such that $n$
also behaves as a thermal particle at temperature $T$,
giving an equilibrium distribution $\rho(n)\propto n^{-d/2}$.

Suppose instead that the chain is subject to forcing which is athermal
with large, non-Gaussian fluctuations at short distances, a reasonable 
assumption given the collisions with the plate.  One can
show that the average force
remains proportional to $-1/n$, but the noise in (\ref{cdyn}) also becomes
athermal.  The effect of this is most easily understood in the
overdamped limit of (\ref{cdyn}).  Then, $n$ executes a biased random
walk with a drift of order $1/n$ \cite{gillis}, with additional random
jumps of varying size due to noise.  As a result, large jumps are
relatively less biased compared with small jumps, and detailed balance
is violated as in the experiment.  At large $n$ this gives biased
diffusion, with the diffusivity determined by the mean-square step
size.  This yields $\rho\propto n^{-\alpha}$, $\alpha \neq d/2$.  For
smaller $n$, the probability distribution is determined only by the
smaller jumps which are more biased, giving a sharpening of the peak in the
distribution at a width of the order of the largest jump size, consistent
with observations.  This behavior is independent of the precise form of the
noise, as confirmed by our numerical simulations of (\ref{cdyn}).

Finally, we consider the dynamics of constraints in linear chains 
instead of rings.  In this case, knots can open at the ends of
the chain.  Consider a linear chain which crosses itself at
one point, the analogue of the figure-8 considered here.  Let the
crossing occur at links $n_1,n_2$, with \hbox{$0<n_1<n_2<N$}.  The
points $n_1,n_2$ describe a random walk, with boundary conditions
$n_1<n_2$, and a bias proportional to $1/(n_2-n_1)$.  For $\alpha>1$,
the two walkers form a bound state, and the time for the knot to open
will behave for large $N$ as for a single random walker.  This
contrasts with the behavior found in a larger acceleration regime, for
which experimental measurements of knot opening times showed a purely
diffusive behavior\cite{bdve}, with negligible entropic interaction
between walkers.  We speculate that the reason for the reduced
interaction in the larger acceleration regime is that the increased
drive takes the system further out of equilibrium.

In conclusion, we have observed a spontaneous tightening of
topological constraints in vertically vibrated granular chains.  The
significance of this phenomenon is that it indicates that the presence
of the constraint merely reduces the chain length by a fixed amount,
rather then leading to an extensive size reduction.  For equilibrium
polymers the tightening arises from entropy. Here, due to the strongly
nonequilibrium drive applied to the system, the bead dynamics is
athermal and the crossing dynamics breaks detailed balance.  However,
the loop-size distribution remains close to equilibrium.  This system
provides further possibilities for experimental examination of the
role of entropic forces in nonequilibrium statistical mechanics.
It is possible to probe fluctuation-dissipation relations
by a quantitative comparison between the forces on the
crossing point and the velocity fluctuations in the beads, namely the
granular temperature.

We thank Charles Reichhardt for useful discussions. This work was
supported by US DOE(W-7405-ENG-36) and by the Canadian NSERC.


\vskip-5mm
\begin{thebibliography}{99}
\bibitem{n} S.~Nachaev, {\sl Statistics of Knots and Entangled
Random Walks} (World Scientific, Singapore, 1996).

\bibitem{fw} H.~L.~Frish and E.~Wasserman, J. Am. Chem. Soc. {\bf 83}, 3789 (1961). 

\bibitem{sw} D.~W.~Sumners and S.~G.~Wittington, J. Phys. A {\bf 21}, 1689 (1989).

\bibitem{sw1} S. Y. Shaw and J. C. Wang, Science {\bf 260}, 533 (1993).

\bibitem{q} S. R. Quake, Phys. Rev. Lett. {\bf 73}, 3317 (1994).

\bibitem{bgwb} E. Ben-Naim, G. S. Grest, T. A. Witten, and A. R. C. Baljon,
Phys. Rev. E {\bf 53} 1806, (1996).

\bibitem{dg} P. G. de Gennes, {\it Scaling Concepts in Polymer Physics}
(Cornell, Ithaca, 1979).

\bibitem{de} M.~Doi and S.~F.~Edwards, {\it The Theory of Polymer
Dynamics} (Clarendon Press, Oxford, 1986).

\bibitem{kovds} V.~Katrich, W.~K.~Olson, A.~Vologodskii,
J.~Dubochet, and A.~Stasiak, Phys. Rev. E {\bf 61}, 5545 (2000).

\bibitem{g} A. Yu. Grosberg, Phys. Rev. Lett. {\bf 85} 3858 (2000).

\bibitem{mhdkk} R.~Metzler, A.~Hanke, P.~G.~Dommersnes, Y.~Kantor, M.~Kardar, 
cond-mat/0110266.

\bibitem{qbc} S.~R.~Quake, H.~Babcock, and S.~Chu, Nature {\bf 388}, 151 (1997). 

\bibitem{k} L. P. Kadanoff, Rev. Mod. Phys. {\bf 71}, 435 (1999).

\bibitem{jnb}H.~Jaeger, S.~Nagel, and R.~P.~Behringer, Rev. Mod.
    Phys. {\bf 68}, 1259 (1996).

\bibitem{ou} J. S. Olafson and J. S. Urbach, Phys. Rev. Lett. {\bf 81},
4369 (1998).

\bibitem{mus} F. Mello, P. B. Umbanhowar, and H. L. Swinney, Phys.
Rev. Lett. {\bf 75}, 3838 (1995).

\bibitem{bdve} E. Ben-Naim, Z. A. Daya, P. Vorobieff, and R. E. Ecke,
Phys. Rev. Lett. {\bf 86}, 1414 (2001).

\bibitem{bsew}A. Belmonte, M.~J.~Shelley, S.~T.~Eldakar, and
C.~H.~Wiggins, Phys. Rev. Lett. {\bf 87}, 114301 (2001).

\bibitem{ds} B.~Duplantier and S.~Saleur, Phys. Rev. Lett. {\bf
57}, 3179 (1986).

\bibitem{gillis} B. D. Hughes, {\it Random Walks and Random Environments,
Vol. 1}, (Clarendon Press, Oxford, 1995).

\end{thebibliography}
\end{document}